\documentclass[sn-mathphys-num]{sn-jnl}

\usepackage{fullpage}
\usepackage{tcolorbox}
\usepackage{subcaption}
\usepackage{todonotes}
\usepackage{graphicx}%
\usepackage{multirow}%
\usepackage{amsmath,amssymb,amsfonts}%
\usepackage{amsthm}%
\usepackage{mathrsfs}%
\usepackage[title]{appendix}%
\usepackage{textcomp}%
\usepackage{manyfoot}%
\usepackage{booktabs}%
\usepackage{algorithm}%
\usepackage{algorithmicx}%
\usepackage{algpseudocode}%
\usepackage[T1]{fontenc}
\usepackage{listings}%
\usepackage{array}
\usepackage[table]{xcolor}
\usepackage{threeparttable}
\usepackage{hyperref}
\hypersetup{
    pdfborder={0 0 1},       
    colorlinks=false,        
    linkbordercolor={0 1 0}, 
    citebordercolor={0 1 0}, 
    urlbordercolor={0 1 0},  
    filebordercolor={0 1 0}, 
    runbordercolor={0 1 0},  
    menubordercolor={0 1 0}  
}
\newcommand{\modified}[1]{\textcolor{black}{#1}}

\begin{document}

\title{Leveraging LLMs to support co-evolution between definitions and instances of textual DSLs: A Systematic Evaluation}

\author*[1]{\fnm{Weixing} \sur{Zhang}}\email{weixing.zhang@kit.edu}

\author[1]{\fnm{Bowen} \sur{Jiang}}\email{bowen.jiang@kit.edu}

\author[2]{\fnm{Yuhong} \sur{Fu}}\email{yuhong.fu@adelaide.edu.au}

\author[1]{\fnm{Anne} \sur{Koziolek}}\email{koziolek@kit.edu}

\author[3]{\fnm{Regina} \sur{Hebig}}\email{regina.hebig@uni-rostock.de}

\author*[4,5]{\fnm{Daniel} \sur{Strüber}}\email{danstru@chalmers.se}

\affil*[1]{\orgdiv{MCSE Group}, 
\orgname{Karlsruhe Institute of Technology}, 
\orgaddress{
\country{Germany}}}

\affil[2]{\orgdiv{Engineering and Information Technology}, 
\orgname{Adelaide University}, 
\orgaddress{
\country{Australia}}}

\affil[3]{\orgdiv{Institute of Computer Science}, 
\orgname{University of Rostock}, 
\orgaddress{
\country{Germany}}}

\affil[4]{\orgdiv{Department of Computer Science and Engineering}, 
\orgname{Chalmers University of Technology and University of Gothenburg}, 
\orgaddress{
\country{Sweden}}}

\affil[5]{\orgdiv{Department of Software Science}, 
\orgname{Radboud University}, 
\orgaddress{
\country{Netherlands}}}

\abstract{
Software languages evolve over time for various reasons, such as the addition of new features. When the language's grammar definition evolves, textual instances that originally conformed to the grammar become outdated. For DSLs in a model-driven engineering context, there exists a plethora of techniques to co-evolve models with the evolving metamodel. However, these techniques are not geared to support DSLs with a textual grammar--- applying them to textual language definitions and instances may lead to the loss of information from the original instances, such as layout information and comments, which are valuable for software comprehension and maintenance. 
\modified{This study systematically evaluates the potential of Large Language Model (LLM)-based solutions in achieving grammar and instance co-evolution for textual DSLs. By applying two advanced language models, Claude Sonnet 4.5 and GPT-5.2, and conducting ten experimental runs per case across ten case languages, we evaluate both the correctness of co-evolved instances and the preservation of human-oriented information such as comments and layout. Our results indicate high performance on small-scale cases ($\geq$94\% precision and recall for instances with fewer than 20 lines requiring modification), but performance degraded with scale: Claude Sonnet 4.5 maintained 85\% recall at 40 lines while GPT-5.2 showed greater sensitivity, failing entirely on the two largest instances. Instance scale also substantially impacts processing efficiency, with Claude's response time increasing nearly 18-fold for the largest case. 
In addition, we observe that grammar evolution complexity and deletion granularity impact performance more than change type alone, and prompt transferability across LLMs is limited. 
These findings identify the conditions under which LLM-based co-evolution succeeds and its scalability limitations, providing practitioners with insights into applicability and researchers with directions for addressing current limitations.}
}



\maketitle

\section{Introduction}
\label{sec:introduction}
Domain-specific languages (DSLs) are specialized languages designed to to describe and solve problems in a specific application domain. As domain knowledge evolves and requirements change, DSLs often need to evolve accordingly \cite{lammel2018software}. For example, features may be added, and existing functionality may be adjusted, leading to a need to update the definition of the DSL, typically specified using grammars and/or meta-models, to introduce new language constructs and modify existing ones.

When the definition of a DSL evolves, existing instances face multiple challenges. First, instances may contain constructs that no longer conform to the new definition, requiring appropriate modification or replacement. Second, the new definition may introduce required language elements, necessitating corresponding additions to existing instances~\cite{martin2006evolution}~\cite{martin2008incremental}. 
These issues lead to a need for dedicated co-evolution approaches \cite{hebig2016approaches}.


The model-driven engineering community has brought forward a plethora of available approaches for metamodel-instance co-evolution \cite{hebig2016approaches}, but these works are generally focused on metamodel-based language definitions, usually in the context of graphical DSLs~\cite{hebig2016approaches}.
\modified{Textual DSLs are widely adopted in practice~\cite{bock2019advantageous}\cite{karu2012textual},}
which can emulate the \textit{look and feel} of familiar general-purpose languages and are easy to integrate with standard developer tools for versioning, differencing, and merging.
Textual DSLs developed in dedicated frameworks like \textit{Xtext}~\cite{bettini2016implementing},  \textit{langium}~\cite{langium}, and \textit{textX}~\cite{dejanovic2017textx}.
On a technical level, textual DSLs are defined through grammars and instantiated by textual instances.

Dedicated approaches to co-evolve textual instances are scarce. One possible way to address the co-evolution of textual instances is by using the available metamodel-based approaches.
To that end, the original instance needs to be parsed into the form of a model and transformed back into textual form after the model is co-evolved.  
However, this approach leads to information loss: during the transformation process between textual instance and model, human-oriented information such as comments and formatting styles in the original instances cannot be retained~\cite{latifaj2021towards}. While this information does not affect the formal semantics of a DSL instance, 
it serves a critical purpose during tasks such as code maintenance, debugging, and understanding design intent~\cite{yang2019survey}.
Hence, there is arguably a need to preserve such information during the co-evolution of instances.

In recent years, Large Language Models (LLMs) have demonstrated exceptional capabilities in code understanding, transformation, and generation~\cite{nam2024using}~\cite{dong2024generalization}~\cite{dao2025learning}. These models do not only perform well at tasks that require an understanding of code structure, but also capture contextual information like comments. As such, they seem particularly well-suited to addressing the co-evolution problem for textual languages.

In this paper, we \modified{systematically evaluate} the use of LLMs to support the co-evolution of grammar definitions and instances for textual DSLs. 
\modified{Given an original grammar, a textual instance conforming to it, and an evolved grammar, we leverage LLMs to automatically analyze the grammar differences and migrate the instance to conform to the evolved grammar while preserving human-oriented information.}
We focus on grammar definitions and instances developed using  Xtext~\cite{bettini2016implementing},  a framework that is rooted in the Eclipse ecosystem and is particularly widely used in the MDE community.
We harness our recently published dataset on Xtext-based language evolution cases retrieved from GitHub~\cite{zhang2024tales, zhang2026development}, which allowed us to identify a collection of real-life cases in which automated support for co-evolution would be desirable. 
\modified{We employ two state-of-the-art Large Language Models—Claude Sonnet 4.5 and GPT-5.2—and conduct ten experimental runs per case language to account for non-determinism in LLM outputs.}
We address and answer the following research questions:\\
\textbf{RQ1: How do 
instance size and grammar change extent 
affect the capability of LLMs to produce correct solutions when co-evolving textual instances?}
\modified{Answering this question allows us to identify the practical boundaries within which LLM-based co-evolution can be applied, and to determine when alternative approaches may be needed.}
\\
\textbf{\modified{RQ2: How does the type of grammar evolution affect LLMs' capability in correctly co-evolving textual instances?}}
\modified{Answering this question allows us to identify which categories of grammar changes LLMs handle well and which pose difficulties, informing practitioners about when to trust LLM-generated solutions.}
\\
\textbf{\modified{RQ3}: How do these factors affect LLMs' capability in preserving human-oriented information during DSL co-evolution?}
\modified{Answering this question allows us to understand under what conditions LLMs can preserve comments and formatting, which is the advantage of LLM-based approaches over traditional model-based co-evolution.}\\
\textbf{\modified{RQ4: To what extent do prompts transfer across different LLM generations?}}
\modified{Answering this question allows us to understand whether a single prompt can work across different LLMs or whether LLM-specific adjustments are necessary.}

\modified{The contribution of this paper is a systematic evaluation of LLM-based co-evolution for textual DSLs across ten real-world case languages. We quantified the correctness of co-evolved instances using precision and recall metrics, characterized grammar evolution types and their impact on co-evolution performance,} measured the preservation of human-oriented information such as comments and formatting, \modified{and examined whether prompt designs transfer across different LLMs. Our findings identify the conditions under which LLM-based co-evolution succeeds and where it faces scalability limitations, providing practitioners with insights on the applicability of this approach and researchers with directions for addressing current limitations.}

\modified{This paper is a significantly extended version of our previous workshop paper~\cite{zhang2025leveraging}, in which we first presented our LLM-based approach and addressed a subset of the research questions stated above. For the present manuscript, we considerably broaden our evaluation scope and introduce new research questions to provide a more in-depth analysis of factors affecting LLM performance in grammar-instance co-evolution. This entailed extensive work on classifying grammar evolution types and analyzing their impact on co-evolution correctness (leading to the new RQ2), investigating whether different LLMs require different prompt designs (RQ4, new), expanding the evaluation metrics to include precision, recall, and normalized percentages, and extending the number of case languages from 7 to 10 to strengthen the generalizability of our findings. Since the LLMs used in our workshop paper (Claude-3.5 and GPT-4o) are now two generations behind the current state-of-the-art, and Claude-3.5 Sonnet has been retired from Anthropic's API, we updated our experiments to use Claude Sonnet 4.5 and GPT-5.2. We retained the same prompting text from the workshop paper to examine whether prompts designed for earlier models remain applicable to their successors, which also provides data for addressing RQ4. Based on the new findings, we also significantly extended our discussion to discuss the scalability limitations of LLM-based co-evolution (significantly extended Section~\ref{sec:scalability}) and the implications for practitioners seeking to apply LLMs in DSL evolution tasks (new Section~\ref{sec:practioner}).}

In subsequent sections, we describe the problem we solve in this paper in Section~\ref{sec:problem} and introduce the background knowledge involved in this paper in Section~\ref{sec:background}. In Section~\ref{sec:methodology}, we present our research methodology and evaluation metrics. Then, Section~\ref{sec:results} presents and analyzes experimental results, with Section~\ref{sec:discussion} discussing the research findings and connections to existing research. Next, we review related research work on DSL evolution and LLM applications in Section~\ref{sec:relatedwork}. Finally, Section~\ref{sec:conclusion} concludes this research.

\section{\modified{Problem Description and Motivation}}
\label{sec:problem}
\begin{figure}[htbp]
\centering
    \includegraphics[width=0.6\linewidth]{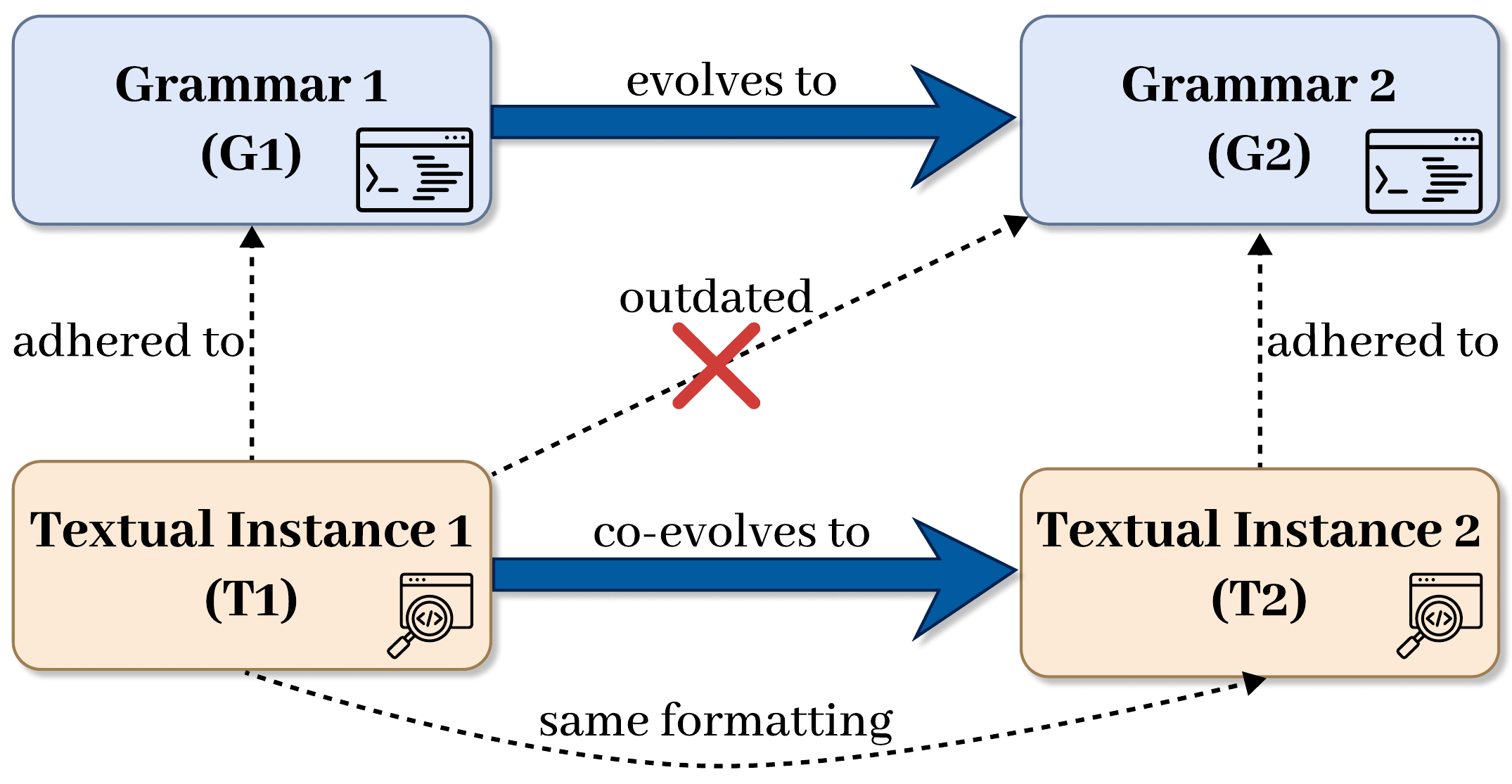}
\caption{When the grammar evolves, textual instances that originally adhered to it 
may no longer conform to it, and so need to be co-evolved to conform to it.
}
\label{fig:problem}
\end{figure}

Xtext is an Eclipse-based framework for developing textual programming languages and domain-specific languages~\cite{Xtext}. 
As core artifacts for defining a language, it involves metamodels, specifying the abstract syntax, and grammars, specifying the context syntax of a language.
As the language evolves, metamodels and grammars need to be systematically co-evolved.
In our previous work \cite{zhang2024supporting}, we have proposed an automated approach that can avoid information loss in grammars when the meta-model is changed and corresponding grammar updates have to be derived.
However, after the grammar evolves, the instances that originally conformed to may no longer conform to it (see Figure~\ref{fig:problem}).
It would be possible to make use of existing techniques for co-evolving models with evolving metamodels.
For that, the textual instance could be parsed using the original grammar to gain the model representation (i.e., in XMI format) that conforms to the original metamodel.
Then existing model migration technologies (such as EMFMigrate~\cite{wagelaar2012translational}) can be used to transform this model into a model that conforms to the new metamodel. Finally, we can transform the migrated model back into a textual instance that conforms to the new grammar.

However, in this process, the human-oriented information such as comments and code formatting in the original text instance is discarded during the first transformation (i.e., when transforming textual instances to model representations). As an example, there is a ``15 minutes tutorial''~\cite{xtext15minstutorial} on the official Xtext website which provides an example called \texttt{Domainmodel}, which includes two versions of grammar and their corresponding instances, where the second version adds five grammar rules. 
Consider a slightly modified version of the instance from the tutorial example with human-oriented information in different places in the instance, shown in  Listing~\ref{lst:dmodel_instance_1}.
Line 10 is empty, which, in the case of entities with many more attributes, is a useful way of grouping them,
The definition of instance \texttt{HasAuthor} has been compressed from  originally four lines to a single line (line 14) by removing whitespace, making the overall instance more compact and easier to overview. 
Line 19 uses comments as a way to discard a part from the instance that might potentially be included again at a later time (outcommenting).
Line 24 contains an additional comment in a style commonly used to add rationale and context to individual statements. 
More subtlely, lines 9 and 11,  use a different type of indentation than the rest of the instance, based on tabs, which could be the result of an ongoing manual review and refactoring.

\begin{lstlisting}[caption={The instance of Domainmodel that conforms to the grammar before the evolution.}, label={lst:dmodel_instance_1}, basicstyle=\footnotesize\ttfamily, frame=single]
/**
 * This is the example before the evolution.
 * This is the header.
 * */
datatype String

/* this is the first comment, added by me */
entity Blog {
	title: String
	
	many posts: Post
}
 
entity HasAuthor { author: String }
 
entity Post extends HasAuthor {
    title: String
    content: String
    //many comment: Comment
    many comments: Comment
}
 
entity Comment extends HasAuthor {
    content: String // this is the second comment, added 2025-01-01
}
\end{lstlisting}

\begin{lstlisting}[caption={The instance of \texttt{DomainModel} that conforms to the evolved grammar and that are generated using traditional model transformation techniques.}, label={lst:dmodel_instance_2_mde}, basicstyle=\footnotesize\ttfamily, frame=single]
datatype String

entity Blog {
    title: String
    many posts: Post
}
 
entity HasAuthor { 
    author: String
}
 
entity Post extends HasAuthor {
    title: String
    content: String
    many comments: Comment
}
 
entity Comment extends HasAuthor {
    content: String
}
\end{lstlisting}

Traditional co-evolution approaches in the MDE sphere focus on co-evolution on the abstract syntax level, that is, the impact of new and changed meta-model classes and relationships to model elements. Concrete syntax information without an abstract syntax counterpart -- that is, human-oriented information such as comments and whitespace -- is not covered and hence, lost in the process.
This is illustrated by Listing~\ref{lst:dmodel_instance_2_mde} showing the outcome of applying such an approach to the exam instance, which leads to the loss of all comments and formatting information.

\section{Background}
\label{sec:background}
\subsection{Language Evolution}
The evolution of programming languages is primarily driven by the need to improve readability, maintainability, and execution efficiency while supporting increasingly complex software systems. Horowitz~\cite{horowitz1984evolution} noted that evolution typically involves grammatical improvements, the introduction of new features, paradigm shifts such as object orientation, and adapting to changing computational requirements. In domain-specific languages (DSLs), language evolution often encompasses changes in syntax, semantics, and implementation methods. According to Cleenewerck et al.~\cite{cleenewerck2005evolution}, DSL evolution may include grammatical extensions (such as adding new keywords), semantic adjustments (like optimizing the execution of specific DSL statements), implementation improvements (such as enhancing compilers), and compatibility maintenance (like providing automatic migration tools). DSL evolution is similar to general-purpose programming languages, but due to its domain-specific nature, evolution is typically constrained by domain requirements, necessitating a balance between extensibility and stability.

\subsection{Xtext-based DSLs}
Xtext is a framework for developing programming languages and domain-specific languages~\cite{Xtext}. Language engineers can define a language using Xtext's powerful grammar language to obtain a complete infrastructure including parsers, connectors, etc. Xtext-based DSL development includes two key artifacts, i.e., grammar and metamodel. Grammar represents the concrete syntax of the language, while metamodel represents the abstract syntax of the language~\cite{wkasowski2023concrete}. Xtext supports two scenarios for developing languages, namely metamodel-driven scenarios and grammar-driven scenarios~\cite{zhang2024supporting}. In the metamodel-driven scenario, language engineers first create a metamodel to represent the domain concepts and their relationships, generate grammar through the metamodel, and then generate the Xtext artifacts with the grammar. Those Xtext artifacts are for generating a textual editor. In the grammar-driven scenario, language engineers directly define the grammar of the language and then generate the infrastructure from the grammar. In this generation, the metamodel will be generated.

\subsection{Large Language Models}
\modified{LLMs are advanced AI models based on the transformer architecture~\cite{vaswani2017attention}, which employs self-attention mechanisms to capture relationships between different parts of the input text. Building on this foundation, models like BERT~\cite{devlin2019bert} introduced bidirectional pre-training techniques that enable deep contextual understanding by jointly conditioning on both left and right context in all layers.}
\modified{LLMs such as ChatGPT and Claude have been widely explored in code-related applications, with several studies reporting promising results in code completion, generation, and transformation tasks~\cite{husein2024large}~\cite{liu2024your}.}
These capabilities can be further enhanced through carefully designed prompts (known as prompt engineering), which guide the models to perform specific text processing and generation tasks~\cite{marvin2023prompt}.

In the context of Model-Driven Engineering (MDE) and Domain-Specific Languages (DSLs), LLMs have 
\modified{been widely explored and applied by}
the research community~\cite{costa2024modelmate}~\cite{rubei2025llm}~\cite{reinhartz2025leveraging}. A search in Google Scholar reveals nearly 10,000 publications related to LLMs in MDE and DSL contexts, indicating substantial research interest in this area. Their potential capability to process both formal structures (like grammar rules) and textual content (like comments) makes them interesting candidates for exploration in MDE tasks.

\section{Methodology}
\label{sec:methodology}
\begin{figure*}[htbp]
\centering
    \includegraphics[width=\linewidth]{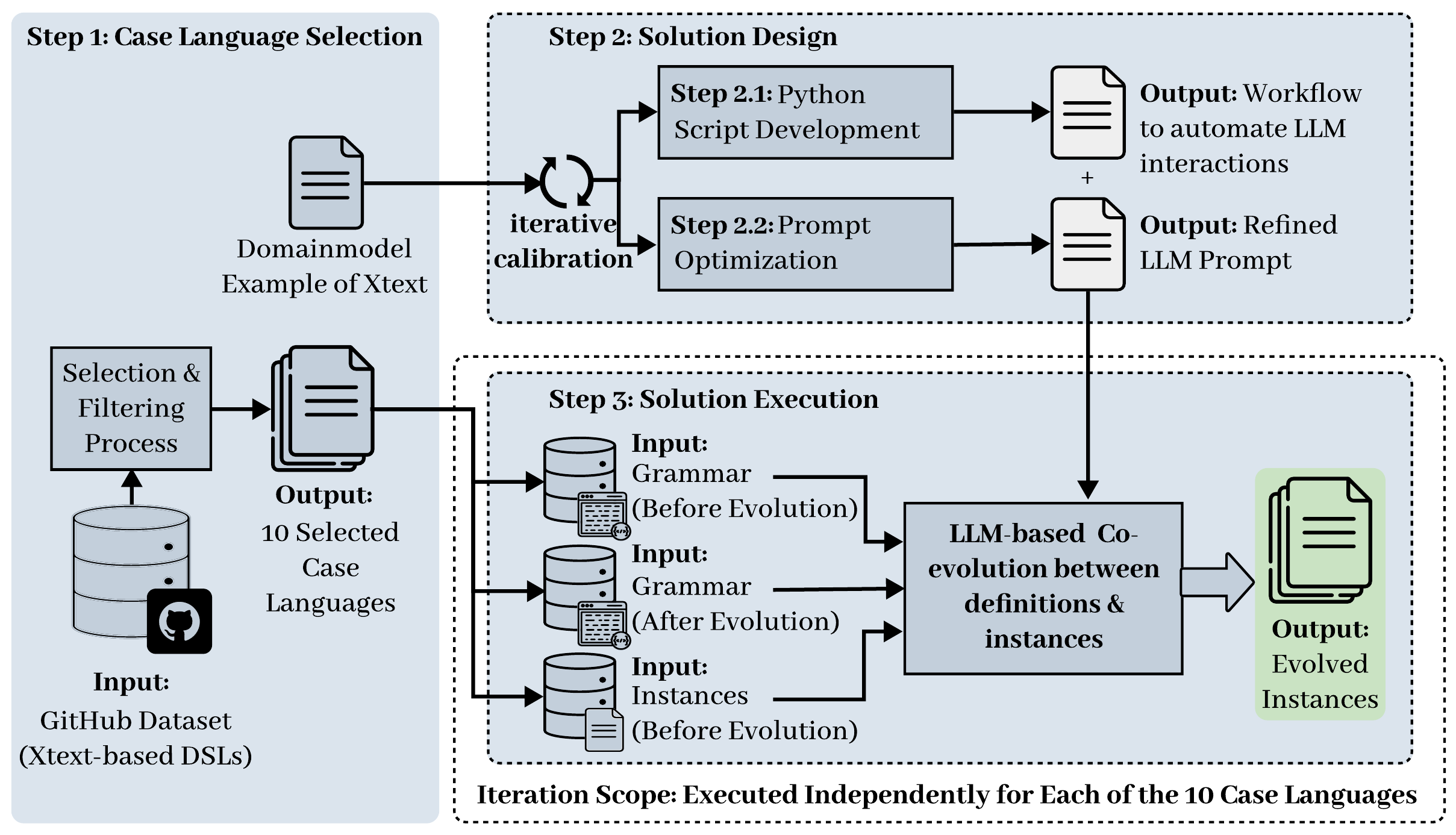}
\caption{Three-step research methodology.}
\label{fig:methodology}
\end{figure*}

\modified{This study was conducted in two phases, each} following the same three-step methodology framework, as depicted in Figure~\ref{fig:methodology}. In the first phase, presented in our workshop paper~\cite{zhang2025leveraging}, we selected seven case languages from an available dataset~\cite{zhang2026development} as evaluation objects (Step 1). We then designed an LLM-based method to co-evolve grammar and instances, including developing Python scripts to automate interactions with LLMs, and iteratively optimizing the prompt text using Claude-3.5 and GPT-4o based on the official Xtext Domainmodel tutorial example~\cite{xtext15minstutorial} (Step 2). Finally, we applied the two solutions implemented with the above LLMs to the seven case languages, evaluating the correctness of co-evolution and the preservation of human-oriented information (Step 3).

\modified{In the second phase, for the present journal extension, we made the following extensions on top of the above. First, we selected three additional case languages (eclipse-typescript-xtext, JSFLibraryGenerator, and majordomo) from the same dataset, expanding the evaluation scope from seven to ten case languages. Second, since Claude-3.5 used in the first phase has been retired from Anthropic's API (as discussed in Section~\ref{sec:introduction}), we updated the LLMs to their current successors, i.e., Claude Sonnet 4.5 and GPT-5.2. Third, we introduced two new research questions, concerning the impact of grammar evolution types on LLM performance (RQ2) and the transferability of prompts across different LLMs (RQ4).}

\modified{One methodological decision requires explanation: we retained the prompt text developed using Claude-3.5 and GPT-4o in the first phase, without re-optimization for the new LLMs or the expanded set of case languages in the second phase. This was a deliberate choice for two reasons. First, it enables us to investigate prompt transferability across LLM generations (RQ4), which would not be possible if the prompt had been re-optimized for the new models. Second, the prompt was designed to address general challenges in LLM-based co-evolution (over-generation of optional elements, omission of mandatory elements, and loss of human-oriented information) and was calibrated on the Domainmodel tutorial example rather than on any of the seven case languages, reducing the risk that it is overfit to the original set of cases. We acknowledge that optimizing the prompt with all ten case languages or with the newer LLMs might yield a different prompt; we return to this point in the discussion of threats to validity (Section~\ref{sec:threats}). The following subsections describe each step of the three-step methodology in detail.}

\subsection{Case Language Selection}
We searched for case languages from in an available dataset~\cite{zhang2024tales}, since this dataset is specifically dedicated to Xtext-based DSLs. We limited our selection to repositories that contain both Xtext files and instance files. From the commit records of the repository, we can see that the Xtext files and instance files contained may contain many commits. For example, in the language \emph{elite-se.xtext.languages.plantuml}, the grammar file ``PlantUML.xtext'' has 63 commits. For each language, we decided to pick a grammar from the commit that is closest to the present and ensure that there must be an instance that complies with the grammar, otherwise we will look for the grammar in earlier commits. The found grammar is the evolved grammar. Then, we continue to look for a grammar that differs from this version in earlier commits. 
Once the grammar versions before and after the evolution step are determined, for each of them, we look for an instance that complies with them. 
We identified \modified{ten} case languages from the dataset. Their basic information is shown in Table~\ref{tab:case_lang}, and the grammars before/after evolution and the instances that comply with this grammar found in their repositories are shown in Table~\ref{tab:grammar_instances_1} \& \ref{tab:grammar_instances_2}.

\begin{table*}[tb]
\scriptsize
\centering
\caption{Case languages and their basic information.}
\label{tab:case_lang}
\resizebox{\textwidth}{!}{
\begin{tabular}{cll}
\toprule
\textbf{Name} & \textbf{Domain} & \textbf{Category} \\
\midrule
xtext-orm~\cite{xtextorm}     & Database   & Data Management and Databases \\
xtext-dnn~\cite{xtextdnn}     & Deep learning    & Artificial Intelligence and Machine Learning \\
smart-dsl~\cite{smartdsl}     & Blockchain     & Security and Networking \\
mongoBeans~\cite{mongoBeans}  & Database access     & Data Management and Databases \\
elite-se.xtext.languages.plantuml~\cite{xtextplantuml} & GPL & Programming Languages \\
isis-script~\cite{isisscript} & Java ISIS applications & Software Development and Engineering \\
CheckerDSL~\cite{CheckerDSL}  & Test cases & Testing and Verification \\
eclipse-typescript-xtext~\cite{typescript-xtext} & GPL & Programming Languages \\
JSFLibraryGenerator~\cite{jsflib} & Web UI / User Interface & Web and Mobile Development \\
majordomo~\cite{majordomo}    & Home Automation & IoT, Embedded Systems, and Hardware \\
\bottomrule
\end{tabular}
}
\end{table*}

In Table~\ref{tab:grammar_instances_1}, \emph{Grammar 1} is a grammar with an earlier commit time, which is regarded as the grammar before the evolution, while \emph{Grammar 2} is a grammar with a later commit time, which is referred to as the evolved grammar. Similarly, in Table~\ref{tab:grammar_instances_2} \emph{Instance 1} is an instance with an earlier commit time, which is the object of LLM evolution operation (we call it the \emph{original instance}), while \emph{Instance 2} is an instance with a later commit time, which conforms to the evolved grammar, but may not be an evolved version of \emph{Instance 1}, because the author of the instance may add or delete content irrelevant to the evolution. We will discuss this situation in the discussion section.

\begin{table*}[tb]
\scriptsize
    \centering
    \caption{Selected grammar and instance versions of the case languages. ``1'' denotes the grammar and instance version before the evolution and ``2'' denotes grammar and instance version after the evolution.}
    \label{tab:grammar_instances_1}
    \begin{threeparttable}
    \begin{tabular}{cllrllr}
      \toprule					
        ~ 	& \multicolumn{3}{c}{\textbf{Grammar 1}} & \multicolumn{3}{c}{\textbf{Grammar 2}} \\
       \cmidrule(llr){2-4}\cmidrule(llr){5-7}
        Name & Date\tnote{1} & ID\tnote{2} & Rules\tnote{3} & Date\tnote{1} & ID\tnote{2} & Rules\tnote{3} \\
        \midrule
        xtext-orm           & 2018-06-21    & f84e2b3   & 15    & 2018-06-23    & 7cb74b2   & 22\\       
        xtext-dnn           & 2016-12-13    & a4912d2   & 31    & 2016-12-17    & 15ccbc0   & 30\\ 
        smart-dsl           & 2023-07-22    & 64259ba   & 8     & 2023-07-31    & 23424ac   & 9 \\ 
        mongoBeans          & 2012-06-04    & 279ecc8   & 9     & 2012-06-06    & 9f8360b   & 9 \\ 
        elite-se.xtext.languages.plantuml & 2020-07-12 & 98f445d & 52 & 2020-07-13 & a41d99f & 55\\
        isis-script         & 2015-07-18    & 2546850   & 15    & 2015-09-05    & f0380e8   & 21\\
        CheckerDSL          & 2015-05-03    & 55911bf   & 42    & 2015-07-27    & 3fa6e6d   & 43 \\
        eclipse-typescript-xtext & 2013-11-16 & 505cc89 & 23    & 2013-11-16    & 58e6e39   & 32 \\
        JSFLibraryGenerator & 2016-03-27    & 44a1b28   & 6     & 2016-09-25    & 20efe92   & 7 \\
        majordomo           & 2013-02-06    & cc01ae0   & 47    & 2013-02-08    & a9967e0   & 47 \\
      \bottomrule
    \end{tabular}
    \begin{tablenotes}		
        \item[1] ``Date'' = the commit date of the grammar file.
        \item[2] ``ID'' = the commit ID of the grammar file's commit.
        \item[3] ``Rules'' = the total number of grammar rules in the grammar.
    \end{tablenotes}
    \end{threeparttable}
\end{table*}

\begin{table*}[tb]
\scriptsize
    \centering
    \caption{Selected grammar and instance versions of the case languages. ``1'' denotes the grammar and instance version before the evolution and ``2'' denotes grammar and instance version after the evolution.}
    \label{tab:grammar_instances_2}
    \begin{threeparttable}
    \begin{tabular}{cllrllr}
      \toprule					
        ~ 	& \multicolumn{3}{c}{\textbf{Instance 1}} & \multicolumn{3}{c}{\textbf{Instance 2}}\\
       \cmidrule(llr){2-4}\cmidrule(llr){5-7}
        Name & Date\tnote{1} & ID\tnote{2} & Lines\tnote{3} & Date\tnote{1} & ID\tnote{2} & Lines\tnote{3} \\
        \midrule
        xtext-orm           & 2018-06-21    & 181dcd7   & 72    & 2018-06-23    & 7cb74b2   & 85\\       
        xtext-dnn           & 2016-12-13    & a4912d2   & 33    & 2016-12-17    & 15ccbc0   & 42 \\ 
        smart-dsl           & 2023-07-22    & 64259ba   & 30    & 2023-07-31    & 23424ac   & 30 \\ 
        mongoBeans          & 2012-06-04    & 279ecc8   & 52    & 2012-06-06    & 9f8360b   & 29 \\ 
        elite-se.xtext.languages.plantuml & 2020-07-12 & 98f445d & 60 & 2020-07-13 & a41d99f & 60 \\
        isis-script         & 2015-07-22    & 1c88416   & 98    & 2015-09-05    & f0380e8   & 68 \\
        CheckerDSL          & 2015-05-03    & 55911bf   & 173   & 2015-07-27    & 3fa6e6d   & 181 \\
        eclipse-typescript-xtext & 2013-11-16 & 505cc89 & 8     & 2013-11-16    & 58e6e39   & 11 \\
        JSFLibraryGenerator & 2016-03-27    & 44a1b28   & 1822  & 2016-09-25    & 20efe92   & 2176 \\
        majordomo           & 2013-02-06    & cc01ae0   & 85    & 2013-02-08    & a9967e0   & 86 \\
      \bottomrule
    \end{tabular}
    \begin{tablenotes}		
        \item[1] ``Date'' = the commit date of the grammar file.
        \item[2] ``ID'' = the commit ID of the grammar file's commit.
        \item[3] ``Lines'' = the total count of lines in the instance, empty lines included.
    \end{tablenotes}
    \end{threeparttable}
\end{table*}

\subsection{\modified{Characterization of Grammar Changes}}
\modified{To answer RQ2, we need to understand what types of changes occurred between \textit{Grammar 1} and \textit{Grammar 2} in each case language. We performed a manual comparison of the two grammar versions for each of the ten cases.}

\modified{For each case, we compared the grammar files line by line and identified the differences between the two versions. We recorded each change and categorized it along two dimensions.}

\modified{The first dimension distinguishes between breaking and non-breaking changes~\cite{hebig2016approaches}. A \textit{breaking change} requires modifications to existing instances for them to conform to the new grammar. For example, if a keyword is renamed from \texttt{`Modifier'} to \texttt{`validator'}, all instances using the old keyword must be updated. A \textit{non-breaking change} -- for example, adding an optional attribute or a new rule that existing instances do not use -- does not require modifications to existing instances. }

\modified{The second dimension describes the operation type of each change. We use four descriptive categories: \textit{Add} (a new rule, attribute, or keyword is introduced), \textit{Delete} (an existing rule, attribute, or keyword is removed), \textit{Rename} (an element keeps its structure but changes its name), and \textit{Modify} (the structure or syntax of an existing element is changed). These categories are not intended as a formal taxonomy but as descriptive labels to characterize the changes we observed.}

\modified{We applied this characterization process to all ten case languages. The results of this analysis are presented in Section~\ref{sec:impact_of_type}, where we examine how these grammar change characteristics relate to LLM performance.}

\subsection{Solution Design}
In step 2, we designed a solution that leverages LLM to enable the co-evolution of grammar and instances in DSL and preserve human-oriented information in the instances during evolution. We will present the design of the solution in this section.

To implement and evaluate this approach, we selected two mainstream LLMs: Claude and ChatGPT, given their demonstrated capabilities in code understanding and generation tasks. The approach takes three inputs: the original grammar, an instance conforming to it (i.e., the instance to be evolved), and the evolved grammar, and contains a prompt we designed. With these inputs, LLMs are expected to analyze the grammar differences and perform instance evolution accordingly. This section first presents the detailed workflow of our approach, followed by the design of the prompting strategy. 

\subsubsection{Python Script Development}
To achieve automated interaction with LLMs, we developed separate Python scripts for Claude Sonnet 4.5 and GPT-5.2 to replace manual operations on LLM product websites such as ChatGPT. These scripts require the configuration of corresponding API keys to access LLM services: Claude Sonnet 4.5 requires Anthropic's API key, while GPT-5.2 requires OpenAI's API key. 
{We applied for API keys from Claude and OpenAI respectively, and configured them in the Python scripts. 
We configure the maximum number of tokens that the model can generate in the current request (i.e., the \texttt{max\_tokens} field) to 
\modified{64000,}
which is a large number for the texts of the files from the ten case languages in this paper (Our consideration is that if we find that it is not large enough, we will increase it).
Although the two scripts call different APIs, they follow the same workflow.

The workflow is as follows: First, the script reads three input files from specified local paths: the original grammar, the evolved grammar, and the instances that conform to the original grammar. 
Considering that our solution does not require explicit training for a specific task, but allows LLM to analyze directly based on the prompt text we provide, the prompting technique we adopt is Zero-shot Chain-of-thoughts prompting~\cite{zhao2023enhancing}.
\modified{We chose zero-shot over few-shot prompting~\cite{brown2020language} for two reasons: (1) grammar evolution varies considerably across cases in change types and complexity, making it difficult to select representative examples that generalize rather than bias the LLM; (2) few-shot examples risk causing the LLM to overfit to the demonstrated evolution style rather than reasoning about the actual grammar differences. We consider a comparison with few-shot prompting as future work.}
In our implementation, we send the text of one file at a time to the LLM, along with an additional sentence of hint text, such as \emph{``Here is the initial version of the grammar (i.e., Grammar 1). Please remember this for future reference''}. The benefit is that this approach can be applied to larger languages in the future, reducing the likelihood that the total amount of text exceeds the LLM context window limit. When sending each text file, we do not require the LLM to analyze it, but only remember it. Only after all the text has been sent to the LLM do we start grammar analysis and instance evolution.
Finally, the script saves the evolved instance generated by the LLM to the local file system.

\modified{The Python scripts we created for this paper, grammars of the ten case languages, instances cloned from GitHub repositories, and instances generated by LLMs are available in the supporting materials~\cite{supplemental2025}.}

\subsubsection{Prompt Optimization}
To obtain a prompt that can effectively guide the LLM to co-evolve instance, we started with an initial prompt that we iteratively refined in the two LLM solutions based on the example \emph{Domainmodel} in the official ``15 minutes tutorial'' of Xtext~\cite{xtext15minstutorial}. In this example, we made some changes to the instance before evolution and the grammar after evolution. The changes made to the instance before evolution have been introduced in Section~\ref{sec:problem}, and we will introduce the changes to the grammar after evolution below. In each iteration, we used the prompt to drive the LLM to evolve the instance and then observe whether there are problems with the output instance, e.g., incorrectly modified elements. If there were problems with the output instance, we adjusted and optimized the prompt according to the problem, and entered the next iteration.

LLMs are generally affected by non-determinism, which we need to account for when evaluating the capability of the resulting approach.
To this end, we repeated the co-evolution.
When the instance was correctly co-evolved, we performed nine more co-evolution runs with the same prompt. Considering the uncertainty of LLM outputs, we decided that when at least six of the ten runs output good instances, we would use this version of the prompt as the final version. The out instance is good if it follows the evolved grammar and retains human-oriented information. The grammar before evolution is shown in Table~\ref{lst:dmodel_grammar_1}, which contains five grammar rules.

The same tutorial provides an evolved grammar. 
This evolution is adding five new grammar rules, but does not involve any changes about the symbols (e.g., ';') in the grammar. To make the evolution changes also reflected in the symbol changes,
we make two modifications to the evolved version, i.e., 1) in the \texttt{Entity} rule, we add commas (`,') to the attribute features to separate it, instead of just spaces; 2) add a semicolon (`;') as a terminator at the end of the \texttt{DataType} rule. In addition, we also deliberately added an optional attribute called \texttt{default} in the grammar rule \texttt{Feature} 
which means that LLMs need to identify more changes in the grammar.
The final evolved grammar is shown in Listing~\ref{lst:dmodel_grammar_2}.

\begin{lstlisting}[caption={The grammar of Domainmodel before the evolution.}, label={lst:dmodel_grammar_1}, basicstyle=\footnotesize\ttfamily, frame=single]
...
 
Domainmodel:
    (elements+=Type)*;
 
Type:
    DataType | Entity;
 
DataType:
    'datatype' name=ID;
 
Entity:
    'entity' name=ID ('extends' superType=[Entity])? '{'
        (features+=Feature)*
    '}';
 
Feature:
    (many?='many')? name=ID ':' type=[Type];
\end{lstlisting}

\begin{lstlisting}[caption={The grammar of Domainmodel after the evolution.}, label={lst:dmodel_grammar_2}, basicstyle=\footnotesize\ttfamily, frame=single]
...
Domainmodel:
    (elements+=AbstractElement)*;
 
PackageDeclaration:
    'package' name=QualifiedName '{'
        (elements+=AbstractElement)*
    '}';
 
AbstractElement:
    PackageDeclaration | Type | Import;
 
QualifiedName:
    ID ('.' ID)*;
 
Import:
    'import' importedNamespace=QualifiedNameWithWildcard;
 
QualifiedNameWithWildcard:
    QualifiedName '.*'?;
 
Type:
    DataType | Entity;
 
DataType:
    'datatype' name=ID ';';
 
Entity:
    'entity' name=ID ('extends' superType=[Entity|QualifiedName])? '{'
        //(features+=Feature)*
        (features+=Feature  (',' features+=Feature)*)?
    '}';
 
Feature:
    (many?='many')? name=ID ':' type=[Type|QualifiedName] ('(' default=ID ')')?;
\end{lstlisting}

The tutorial also provides an instance that conforms to the grammar before the evolution.
But as we mentioned in Section~\ref{sec:problem}, we added comments and format information to the instance before evolution (as shown in Listing~\ref{lst:dmodel_instance_1}) to evaluate whether the solution can preserve human-oriented information during co-evolution. We added four comments,
one of which is changed from a normal instance line. 
In addition, we added an empty line and two tabs at different locations and put a multi-line code block into one line. Under the guidance of the prompting text, LLMs are expected to correctly identify this formatting information and replay it in the evolved instance.

We began with an initial prompting text that simply outlined the co-evolution task:\\
\begin{tcolorbox}[colback=white!100!black, colframe=white!50!black, arc=3mm, left=0.5em, right=0.5em, top=0.5em, bottom=0.5em]
\textbf{Initial prompt}:\\
grammar\_1 is the initial grammar of the DSL. We evolved it to get grammar\_2. instance\_1 was originally a text instance that followed grammar\_1. Now I want you to analyze the differences between the two versions of the grammar and, based on this difference, modify instance\_1 and get instance\_2, which will follow grammar\_2.
\end{tcolorbox}


\subsection{\modified{Evaluation Metrics}}
\label{sec:eval_metrics}
In Step 3, we conducted solution execution and result evaluation. During the execution, we applied two Python scripts containing the final version of prompting text (one based on GPT-5.2 and the other on Claude Sonnet 4.5) to ten case languages. 
\modified{For each case language, we repeat the co-evolution ten times to account for non-determinism in LLM outputs, resulting in ten evolved instances per LLM per case language. We then established a set of metrics to measure different aspects of the evolved instances. These metrics are organized into three categories: correctness metrics, human-oriented information preservation metrics, and efficiency metrics.}
\modified{All metrics except response time were calculated through manual inspection of the evolved instances, as described below.}

\subsubsection{\modified{Correctness Metrics}}
\modified{These metrics assess how correctly the LLM performs the co-evolution task.}

\textbf{\modified{Basic Counts:}}
\begin{itemize}
    \item \modified{\textit{\#LineReq}: Count of lines in instance 1 that require modification to conform to the evolved grammar.}
    \item \textit{\#LineErr}: Count of lines with grammar errors in the evolved instance.
    \item \textit{\#LineEvl}: Count of lines in instance 1 that are \modified{evolved}.
    \item \textit{\#LineEvlWrg}: Count of lines of instance 1 that are missing (or incorrectly evolved) in the evolved instance.
\end{itemize}

\textbf{Derived Ratios:}
\modified{To enable comparison across instances of different sizes, we compute the following normalized metrics:}
\modified{\begin{itemize}
    \item \textbf{Evolution Precision}: The proportion of LLM-modified lines that are correct.
    \begin{equation*}
        \text{Precision} = \frac{\#LineEvl - \#LineEvlWrg}{\#LineEvl}
    \end{equation*}
    \item \textbf{Evolution Recall}: The proportion of lines requiring evolution that are correctly evolved.
    \begin{equation*}
        \text{Recall} = \frac{\#LineEvl - \#LineEvlWrg}{\#LineReq}
    \end{equation*}
    \item \textbf{Error Rate}: The proportion of lines in the evolved instance that contain grammar errors.
    \begin{equation*}
        \text{ErrorRate} = \frac{\#LineErr}{\text{Total lines in evolved instance}}
    \end{equation*}
\end{itemize}}

\subsubsection{\modified{Human-Oriented Information Preservation Metrics}}
\modified{These metrics assess the LLM's capability to preserve comments and formatting information during co-evolution.}

\textbf{\modified{Basic Counts:}}
\begin{itemize}
    \item \textit{\#LineCmtLost}: Count of lines of instance 1 with comments that are lost.
    \item \textit{\#LineCmtSave}: Count of lines of instance 1 with comments that are maintained.
    \item \textit{\#LineFmtLost}: Count of lines of instance 1 with format information that is lost.
    \item \textit{\#LineFmtSave}: Count of lines of instance 1 with format information that is maintained.
\end{itemize}

\textbf{Derived Ratios:}
\modified{\begin{itemize}
    \item \textbf{Comment Retention Rate}: The proportion of comment lines that are preserved.
    \begin{equation*}
        \text{\#CmtRet} = \frac{\#LineCmtSave}{\#LineCmtSave + \#LineCmtLost}
    \end{equation*}
    \item \textbf{Format Retention Rate}: The proportion of formatting lines that are preserved.
    \begin{equation*}
        \text{\#FmtRet} = \frac{\#LineFmtSave}{\#LineFmtSave + \#LineFmtLost}
    \end{equation*}
\end{itemize}}

\modified{These rates are only computed for instances that contain comments or formatting information. For instances without comments, the comment retention rate is not applicable (N/A).}

\subsubsection{\modified{Efficiency Metrics}}
\modified{To provide practical guidance for practitioners, we also measure the time required for LLM-based co-evolution.}
\modified{\begin{itemize}
    \item \textbf{Response Time:} The wall-clock time (in seconds) from sending the prompt to receiving the complete evolved instance from the LLM API.
\end{itemize}}
\modified{We report the average response time across the ten runs for each case language. While we do not conduct a formal comparison with manual co-evolution time (which would require a controlled user study), we discuss the practical implications of the observed response times in Section~\ref{sec:discussion}.}


\section{Results}
\label{sec:results}
In this section, we present two aspects of our results: 1) the final version of the prompting text we obtained; 2) the results obtained after applying our LLM-based solutions to the ten case languages, i.e., the instances evolved by the LLMs according to the evolution of the grammar, and our evaluation.

\subsection{Finalization of Prompting Text}
\modified{Following the method described in Step 2, we refined the prompt through two iterations using the \emph{Domainmodel} example. The initial prompt produced syntactically valid but suboptimal results, revealing three key challenges in LLM-based co-evolution: (1) LLMs tend to over-generate by instantiating newly added optional elements, (2) LLMs may under-generate by omitting mandatory syntactic elements such as delimiters, and (3) LLMs do not inherently prioritize preserving human-oriented information. Based on these observations, we formulated three corresponding instructions that address these general challenges rather than case-specific errors. We verified the resulting prompt through ten runs on the \emph{Domainmodel} example with both LLMs—when at least six of the ten runs produced correct results, we adopted it as the final version.}
\modified{This prompting text was developed in our workshop paper [13] using Claude-3.5 and GPT-4o. For this study, we applied the same prompt to Claude Sonnet 4.5 and GPT-5.2. This design choice allows us to examine whether prompts designed for earlier model versions remain applicable to their successors, addressing RQ4 on prompt transferability across LLM generations.}

\modified{Through this prompt,} we obtained instances from the LLMs (i.e., GPT-5.2 and Claude Sonnet 4.5) that complied with the evolved grammar. We verified this version of the prompt text 10 times for both LLM solutions, and in most of these ten times, the instances obtained complied with the evolved grammar. Therefore, we decided to adopt it as the final version of the prompting text, as follows:
\begin{tcolorbox}[colback=white!100!black, colframe=white!50!black, arc=3mm, left=0.5em, right=0.5em, top=0.5em, bottom=0.5em]
\textbf{Final prompt}:\\
grammar\_1 is the initial grammar of the DSL. We evolved it to get grammar\_2. instance\_1 was originally a text instance that followed grammar\_1. Now I want you to analyze the differences between the two versions of the grammar and, based on this difference, modify instance\_1 and get instance\_2, which will follow grammar\_2. Please address the following things:\\
1.	When evolving the instance, please do not omit any mandatory elements, such as characters enclosed by single quotes. \\
2.	If grammar\_2 adds a new grammar rule or a new attribute that is optional or in an ``OR'' relationship (i.e., |), then please do not instantiate it.\\
3.	Do not miss or add any auxiliary information in the instance, e.g., comments, formats (white space, indents, tabs, empty lines, etc.).
\end{tcolorbox}

Compared to the first version, the final version of the prompting text explicitly adds three items that the LLM needs to do. This is based on the problems we encountered in the process of optimizing the prompting text. The first is added because the LLM ignored the symbol `,', which is a mandatory element. The second  is added because LLM would actively instantiate the grammar rule ``PackageDeclaration'' which is a newly added optional rule. The third item is added because LLM partially ignored comments.


\subsection{Correctness Evaluation \modified{(RQ1)}}
\label{sec:model_ele_evl}

For each case language, we executed the co-evolution task ten times with each LLM and computed the average values of all metrics. The ``Average'' rows of Tables~\ref{tab:correctness} shows the average across all \modified{ten} case languages.
\begin{table*}[htbp]
\centering
\caption{Correctness metrics for Claude Sonnet 4.5 and GPT-5.2. Each row represents the average of ten runs.}
\label{tab:correctness}
\begin{threeparttable}
\setlength{\tabcolsep}{4pt}
\renewcommand{\arraystretch}{1.1}
\resizebox{\textwidth}{!}{
\begin{tabular}{l l r r r r r r r}
\toprule
LLM & DSL & \#LineReq & \#LineErr & \#LineEvl & \#LineEvlWrg & Precision (\%) & Recall (\%) & ErrorRate (\%) \\
\midrule
\multirow{8}{*}{Claude}
& xtext-orm   & 19 & 0   & 19   & 0   & 100   & 100   & 0    \\
& xtext-dnn   & 15 & 0   & 15   & 0.8 & 94.67 & 94.67 & 0    \\
& smart-dsl   & 3  & 0   & 3    & 0   & 100   & 100   & 0    \\
& mongoBeans  & 4  & 0   & 4    & 0   & 100   & 100   & 0    \\
& plantuml\tnote{1} & 2 & 0 & 5   & 0   & 100   & 100   & 0 \\
& isis-script & 40 & 5.6 & 36.8 & 2.8 & 92.39 & 85 & 4.86 \\
& CheckerDSL  & 8  & 0   & 8    & 0   & 100   & 100   & 0 \\
& typescript\tnote{3} & 0 & 0 & 0 & 0 & N/A & N/A & 0 \\
& JSFLibraryGenerator & 0 & 0 & 2 & 0 & 100 & N/A & 0 \\
& majordomo & 7 & 0 & 7 & 0 & 100 & 100 & 0 \\
\rowcolor{gray!15}
& \textbf{Average} & \textbf{9.8} & \textbf{0.56} & \textbf{9.98} & \textbf{0.36} & \textbf{98.56} & \textbf{97.5} & \textbf{0.49} \\
\midrule
\multirow{8}{*}{GPT}
& xtext-orm   & 19 & 1.2 & 15.8 & 3   & 78    & 67.37 & 4.20 \\
& xtext-dnn   & 15 & 0   & 15   & 0   & 100   & 100   & 0 \\
& smart-dsl   & 3  & 0   & 3    & 0   & 100   & 100   & 0 \\
& mongoBeans  & 4  & 0   & 4    & 0   & 100   & 100   & 0 \\
& plantuml\tnote{1} & 2 & 0 & 2.9 & 0 & 100 & 100 & 0 \\
& isis-script & 40 & 7.4 & 43.4 & 15.4 & 64.52 & 64.3 & 7.60 \\
& CheckerDSL\tnote{2} & 8 & 173 & 0 & 0 & 0 & 0 & 0 \\
& typescript\tnote{3} & 0 & 0 & 0 & 0 & N/A & N/A & 0 \\
& JSFLibraryGenerator & 0 & 571.7 & 1033.2 & 1033.2 & 0 & N/A & 4.20 \\
& majordomo & 7 & 0 & 7 & 0 & 100 & 100 & 0 \\
\rowcolor{gray!15}
& \textbf{Average} & \textbf{9.8} & \textbf{75.33} & \textbf{112.43} & \textbf{105.16} & \textbf{71.44} & \textbf{78.96} & \textbf{11.60} \\
\bottomrule
\end{tabular}
}
\begin{tablenotes}[flushleft]
\begin{minipage}{\textwidth}
\raggedright\noindent
\item[1] Full name: elite-se.xtext.languages.plantuml.
\item[2] GPT-5.2 failed to generate valid evolved instances for CheckerDSL in all ten runs.
\item[3] Full name: eclipse-typescript-xtext. 
\end{minipage}
\end{tablenotes}
\end{threeparttable}
\end{table*}


The upper portion of Table~\ref{tab:correctness} presents the correctness metrics for Claude Sonnet 4.5. \modified{Seven of the ten} case languages (xtext-orm, smart-dsl, mongoBeans, plantuml, CheckerDSL, \modified{and majordomo}) achieved 100\% precision and 100\% recall with zero grammar errors across all runs. xtext-dnn showed minor deviations with 94.67\% precision and recall, caused by an average of 0.8 incorrectly evolved lines per run. 
isis-script exhibited the lowest performance among cases with moderate size, with 92.39\% precision, 85\% recall, and a 4.86\% error rate. This case language has 40 lines requiring modification and involves 21 grammar changes (as shown in Table~\ref{tab:grammar-changes}). 
\modified{JSFLibraryGenerator, with a substantially larger Instance 1 (1822 lines), represents an edge case where the instance requires no modifications to conform to the evolved grammar (\#LineReq = 0). Claude Sonnet 4.5 correctly identified this, making only minor adjustments (average two lines per run) with 100\% precision. The overall average across all case languages shows 98.56\% precision, 97.5\% recall, and 0.49\% error rate.}

The lower portion of Table~\ref{tab:correctness} presents the correctness metrics for GPT-5.2. \modified{Five} case languages (xtext-dnn, smart-dsl, mongoBeans, plantuml, \modified{and majordomo}) achieved 100\% precision and recall with zero errors. 
\modified{However, GPT-5.2 failed to generate valid evolved instances for both CheckerDSL and JSFLibraryGenerator in all ten runs, resulting in 0\% precision and 0\% recall for these two cases. 
}
isis-script exhibited relatively lower performance among successful cases, with 64.52\% precision and 64.3\% recall. 
The overall average shows 71.44\% precision, 78.8\% recall, and 11.60\% error rate, though \modified{the failures of CheckerDSL and JSFLibraryGenerator contribute} substantially to the elevated error rate.

\modified{Comparing the two LLMs, Claude Sonnet 4.5 outperformed GPT-5.2 across all aggregate metrics. The precision difference is approximately 27 percentage points (98.56\% vs. 71.44\%), and the recall difference is approximately 18.5 percentage points (97.5\% vs. 78.96\%). Both LLMs performed well on the five smaller case languages (xtext-dnn, smart-dsl, mongoBeans, plantuml, and majordomo), which have fewer than 20 lines requiring modification. Performance degradation occurred primarily in cases with larger instances or more extensive grammar changes. isis-script (40 lines requiring modification, 21 grammar changes) posed difficulties for both LLMs. CheckerDSL (173 lines) caused complete failure for GPT-5.2 but was handled successfully by Claude Sonnet 4.5. JSFLibraryGenerator represents an edge case where no modifications were required; Claude Sonnet 4.5 correctly identified this  while did not.}


\begin{tcolorbox}[colback=white!95!black, colframe=white!50!black, arc=3mm, left=0.5em, right=0.5em, top=0.5em, bottom=0.5em]
\textbf{Answer to \modified{RQ1:}} 
\modified{The capability of LLMs to produce correct co-evolution solutions varies with problem scale. For smaller cases in our dataset, both Claude Sonnet 4.5 and GPT-5.2 achieved high correctness, with precision and recall at or near 100\%. For larger or more complex cases, performance differences emerged: Claude Sonnet 4.5 maintained 85\% recall for isis-script (40 lines requiring modification) and achieved 100\% precision for JSFLibraryGenerator. GPT-5.2 showed greater sensitivity to scale and complexity, dropping to 64.3\% recall for isis-script and failing entirely for both CheckerDSL and JSFLibraryGenerator.}
\end{tcolorbox}

\subsection{\modified{Impact of Grammar Evolution Types (RQ2)}}
\label{sec:impact_of_type}
To answer RQ2, we first characterized the grammar changes in our case languages, then examined how these characteristics relate to LLM performance in co-evolving instances.

\subsubsection{\modified{Grammar Change Characteristics}}
\label{sec:grammar_change_char}
\modified{We analyzed the grammar changes between Grammar 1 and Grammar 2 for each of the ten case languages. Table~\ref{tab:grammar-changes} summarizes our findings.}

\begin{table*}[htbp]
\centering
\scriptsize
\caption{Grammar change characteristics of the ten case languages}
\label{tab:grammar-changes}
\begin{threeparttable}
\resizebox{\textwidth}{!}{
\begin{tabular}{lcccc}
\toprule
Case Language & Total Changes & Breaking & Non-breaking & Primary Operation Types \\
\midrule
CheckerDSL & 5 & 4 & 1 & Modify \\
plantuml\tnote{1} & 20 & 4 & 16 & Modify, Add \\
xtext-orm & 13 & 2 & 11 & Add \\
mongoBeans & 2 & 1 & 1 & Delete, Add \\
smart-dsl & 6 & 2 & 4 & Rename \\
xtext-dnn & 15 & 14 & 1 & Rename \\
isis-script & 21 & 6 & 15 & Add, Modify, Delete \\
typescript\tnote{2} & 16 & 0 & 16 & Add, Modify \\
JSFLibraryGenerator & 11 & 1 & 10 & Add \\
majordomo & 11 & 3 & 8 & Modify, Rename \\
\bottomrule
\end{tabular}
}
\begin{tablenotes}
\item[1] Full name: elite-se.xtext.languages.plantuml.
\item[2] Full name: eclipse-typescript-xtext.
\end{tablenotes}
\end{threeparttable}
\end{table*}

\modified{The ten cases exhibit different evolution characteristics. The number of changes ranges from 2 (mongoBeans) to 21 (isis-script). The proportion of breaking changes varies from 15\% (xtext-orm) to 93\% (xtext-dnn).
mongoBeans has the fewest changes: the keyword \texttt{`mongobean'} was removed, and an optional inline type definition was added. In contrast, isis-script has the most changes, involving the removal of the \texttt{repository} concept, restructuring of \texttt{action} and \texttt{parameter} syntax, and addition of a new \texttt{collection} concept.
xtext-dnn stands out with the highest proportion of breaking changes. Its evolution consists of a systematic renaming of keywords from \texttt{kebab-case} to \texttt{camelCase} (e.g., \texttt{image-size} to \texttt{imageSize}), affecting 14 locations in the grammar.
smart-dsl represents a concept renaming case, where \texttt{Modifier} was renamed to \texttt{validator} throughout the grammar, including the keyword, rule name, and references.
xtext-orm and isis-script are characterized by additions. xtext-orm introduced a mandatory \texttt{Configuration} block and a directionality marker for entity fields. isis-script added collection support and restructured how actions and parameters are defined.
CheckerDSL and PlantUML both have modifications as their primary operation type, but differ in breaking change proportion. CheckerDSL has four out of five changes as breaking, while plantuml has only four out of 20.}

\subsubsection{\modified{Relationship between Grammar Changes and LLM Performance}}

We compared the grammar change characteristics in Table~\ref{tab:grammar-changes} with the correctness data in Tables~\ref{tab:correctness}, and analyzed the observed phenomena through specific evolution cases.

\paragraph{\modified{Statistical Observations}}
\modified{The number of changes does not show a linear relationship with LLM performance. xtext-dnn has 15 changes, and both LLMs performed well (100\% precision and recall). isis-script has 21 changes, with degraded performance (Claude-Sonnet-4.5 86.67\% recall, GPT-5.2 63\% recall). mongoBeans (two changes) and smart-dsl (six changes) both achieved 100\% correctness, though this may reflect their smaller overall scale. The proportion of breaking changes also cannot directly predict performance. xtext-dnn has the highest breaking change proportion (93\%), yet both LLMs handled it well. isis-script has a lower proportion (29\%), yet performance was worse.}

\paragraph{\modified{Context Dependency of Delete Operations}}
\modified{In isis-script, the IsisRepository rule was deleted in grammar 2. The correct evolution should remove the corresponding repository fragment from the instance (Listing~\ref{lst:isis_repo_original} shows the original  \texttt{repository} block, which should be removed). Both LLMs failed to execute this deletion correctly in multiple runs, and instead attempted to ``evolve'' its internal content (Listing~\ref{lst:isis_repo_claude} shows an incorrect example made by Claude).}

\begin{lstlisting}[caption={The original \texttt{repository} block in \textit{instance 1} of ``isis-script''.}, label={lst:isis_repo_original}, basicstyle=\footnotesize\ttfamily, frame=single]
    ...
    @DomainServiceLayout(menuOrder = "10")
	repository {

    	@Action(semantics = SemanticsOf.SAFE)
    	@ActionLayout(bookmarking = BookmarkPolicy.AS_ROOT)
		@MemberOrder(sequence = "1")
		action listAll() {
			container.allInstances(SimpleObject)
		}
		...
	}
    ...
\end{lstlisting}

\begin{lstlisting}[caption={The \textit{instance 2} output by Claude shows its error evolution on the \texttt{repository} block of ``isis-script''.}, label={lst:isis_repo_claude}, basicstyle=\footnotesize\ttfamily, frame=single]
    ...
    @DomainServiceLayout(menuOrder = "10")
	service repository {

    	@Action(semantics = SemanticsOf.SAFE)
    	@ActionLayout(bookmarking = BookmarkPolicy.AS_ROOT)
		@MemberOrder(sequence = "1")
		action void listAll {
			body {
				container.allInstances(SimpleObject)
			}
		}
        ...
	}
    ...
\end{lstlisting}

\modified{In xtext-dnn, the \texttt{type} attribute in \texttt{BranchBody} was deleted (see Listing~\ref{lst:branchbody_original} and Listing~\ref{lst:branchbody_evolved}). Both LLMs correctly identified and removed this attribute (see Listing~\ref{lst:branch_original} and Listing~\ref{lst:branch_evolved}).}

\begin{lstlisting}[caption={Grammar rule \texttt{BranchBody} in \textit{grammar 1} of ``xtext-dnn''.}, label={lst:branchbody_original}, basicstyle=\footnotesize\ttfamily, frame=single]
    BranchBody:
        'type' REFERENCE type=BranchType &
        'operation' REFERENCE operation=OperationType
    ;
\end{lstlisting}

\begin{lstlisting}[caption={Grammar rule \texttt{BranchBody} in \textit{grammar 2} of ``xtext-dnn''.}, label={lst:branchbody_evolved}, basicstyle=\footnotesize\ttfamily, frame=single]
    BranchBody:
        'eltwiseOperation' REFERENCE operation=OperationType
    ;
\end{lstlisting}

\begin{lstlisting}[caption={Instance fragment \texttt{branch} in \textit{instance 1} of ``xtext-dnn''.}, label={lst:branch_original}, basicstyle=\footnotesize\ttfamily, frame=single]
    branch (name: "b1" in:"c3" out:32)     {
		operation -> PROD
		type -> residual
        ...
	}
\end{lstlisting}

\begin{lstlisting}[caption={Instance fragment \texttt{branch} in Claude-volved instance of ``xtext-dnn''.}, label={lst:branch_evolved}, basicstyle=\footnotesize\ttfamily, frame=single]
    branch (name: "b1" in:"c3" out:32)     {
		eltwiseOperation -> PROD 
        ...
	}
\end{lstlisting}

\modified{The comparison reveals: xtext-dnn's deletion is a single attribute, while isis-script's deletion is a complete rule containing multiple nested elements. This indicates that LLMs are more reliable when handling attribute-level deletions, but prone to errors when handling rule-level deletions (especially those involving nested structures).}

\paragraph{\modified{Variations within the Same Language}}
\modified{Another change in isis-script showed different behavior. The attribute \texttt{returnType} of \texttt{IsisAction} changed from optional to mandatory. Both LLMs correctly added the return type in almost all runs. Within the same language, \textit{Delete} operations failed while \textit{Modify} operations succeeded, demonstrating that change type alone cannot predict performance.}

\paragraph{\modified{Impact of Overall Complexity}}
\modified{Comparing isis-script and xtext-dnn: both contain Delete operations but with different results. xtext-dnn's 15 changes are mainly systematic keyword renaming (kebab-case to camelCase), with limited structural changes. isis-script's 21 changes include removing the \texttt{repository} concept, restructuring \texttt{action} and \texttt{parameter} syntax, and adding \texttt{collection} concepts, with mixed types and big structural adjustments.
When grammar evolution involves a mix of multiple change types and a lot of structural adjustments, LLMs are more prone to errors on individual changes—even when these change types can be handled correctly in simpler contexts.}

\begin{tcolorbox}[colback=white!95!black, colframe=white!50!black, arc=3mm, left=0.5em, right=0.5em, top=0.5em, bottom=0.5em]
\textbf{\modified{Answer to RQ2:}} 
\modified{Grammar evolution type alone cannot predict LLM performance, e.g., \textit{Delete}-type changes succeeded in xtext-dnn but failed in isis-script. Our case analysis reveals that LLM performance relates to the overall complexity of grammar evolution—when involving a mix of multiple change types and substantial structural adjustments, LLMs are more prone to errors. Deletion granularity is a factor: attribute-level deletions are easier to handle than rule-level deletions (especially those involving nested structures). Future research should focus on the overall complexity of grammar evolution rather than solely on change type classification.}
\end{tcolorbox}

\subsection{\modified{Human-Oriented Information Preservation (RQ3)}}

As described in~\ref{sec:eval_metrics}, we executed the co-evolution task ten times for each case language with the LLM-based approaches. Our data on human-oriented information retention were also averaged from these ten runs, as shown in Tables~\ref{tab:auxiliary}. Note that \modified{five} case languages (xtext-dnn, smart-dsl, plantuml, isis-script, \modified{and typescript}) contain no comments in their \textit{Instance 1}, so comment retention rates are not applicable for these cases.

\begin{table*}[htbp]
\centering
\caption{Human-oriented information preservation metrics for Claude Sonnet 4.5 and GPT-5.2. Each row represents the average of ten runs.}
\label{tab:auxiliary}
\begin{threeparttable}
\setlength{\tabcolsep}{4pt}
\renewcommand{\arraystretch}{1.1}
\resizebox{\textwidth}{!}{
\begin{tabular}{l l r r r r r r}
\toprule
LLM & DSL & \#LineCmtLost & \#LineCmtSave & CmtRet (\%) & \#LineFmtLost & \#LineFmtSave & FmtRet (\%) \\
\midrule
\multirow{8}{*}{Claude}
& xtext-orm         & 0     & 1     & 100 & 0   & 72   & 100 \\
& xtext-dnn         & N/A   & N/A   & N/A\tnote{1} & 0.8 & 32.2 & 97.58 \\
& smart-dsl         & N/A   & N/A   & N/A\tnote{1} & 0   & 30   & 100 \\
& mongoBeans        & 0     & 6     & 100 & 0   & 52   & 100 \\
& plantuml\tnote{2} & N/A   & N/A   & N/A\tnote{1} & 0   & 60   & 100 \\
& isis-script       & N/A   & N/A   & N/A\tnote{1} & 28.3 & 69.7 & 71.12 \\
& CheckerDSL        & 0   & 26    & 100 & 0   & 173  & 100 \\
& typescript\tnote{4} & 0   & 1     & 100 & 0.2 & 7.8 & 97.50 \\
& JSFLibraryGenerator & N/A & N/A   & N/A\tnote{1} & 0 & 1822 & 100 \\
& majordomo         & 0     & 24    & 100 & 0 & 85 & 100 \\
\rowcolor{gray!15}
& \textbf{Average} & \textbf{0} & \textbf{6.44} & \textbf{100} & \textbf{2.93} & \textbf{240.37} & \textbf{96.60} \\
\midrule
\multirow{8}{*}{GPT}
& xtext-orm         & 0     & 1     & 100           & 0   & 72   & 100 \\
& xtext-dnn         & N/A   & N/A   & N/A\tnote{1}  & 0   & 33   & 100 \\
& smart-dsl         & N/A   & N/A   & N/A\tnote{1}  & 0   & 30   & 100 \\
& mongoBeans        & 0     & 6     & 100           & 0   & 52   & 100 \\
& plantuml\tnote{2} & N/A   & N/A   & N/A\tnote{1}  & 0.7 & 59.3 & 98.83 \\
& isis-script       & N/A   & N/A   & N/A\tnote{1}  & 40.6 & 57.4 & 58.57 \\
& CheckerDSL\tnote{3} & 26  & 0     & 0             & 173 & 0 & 0 \\
& typescript\tnote{4} & 0   & 1     & 100           & 0.8 & 7.2 & 90 \\
& JSFLibraryGenerator & N/A & N/A   & N/A\tnote{1}  & 1018.6 & 803.4 & 44.09 \\
& majordomo         & 0     & 24    & 100           & 0 & 85 & 100 \\
\rowcolor{gray!15}
& \textbf{Average} & \textbf{2.89} & \textbf{3.56} & \textbf{80.00} & \textbf{123.37} & \textbf{119.93} & \textbf{79.10} \\
\bottomrule
\end{tabular}
}
\begin{tablenotes}[flushleft]
\begin{minipage}{\textwidth}
\raggedright\noindent
\item[1] Instance 1 contains no comments; CmtRet is not applicable.
\item[2] Full name: elite-se.xtext.languages.plantuml.
\item[3] GPT-5.2 generated empty files in all ten runs for CheckerDSL, resulting in the complete loss of human-oriented information.
\item[4] Full name: eclipse-typescript-xtext. 
\end{minipage}
\end{tablenotes}
\end{threeparttable}
\end{table*}


\modified{The upper portion of Table~\ref{tab:auxiliary} presents the human-oriented information preservation metrics for Claude Sonnet 4.5. For comment preservation, Claude Sonnet 4.5 retained all comments across all runs for the five case languages that contain comments (xtext-orm, mongoBeans, CheckerDSL, eclipse-typescript-xtext, and majordomo), achieving 100\% comment retention rate with zero comments lost. For format preservation, seven of the ten case languages achieved 100\% format retention rate. Both xtext-dnn and eclipse-typescript-xtext showed minor format loss with 97.58\% and 97.50\% retention respectively. isis-script exhibited lower format retention at 71.12\%, with an average of 28.3 formatting lines lost per run. JSFLibraryGenerator maintained 100\% format retention, correctly preserving all 1822 lines of formatting information. The overall average across all case languages shows 100\% comment retention and 96.60\% format retention.}


\modified{The lower portion of Table~\ref{tab:auxiliary} presents the human-oriented information preservation metrics for GPT-5.2. For comment preservation, GPT-5.2 achieved 100\% retention for xtext-orm, mongoBeans, eclipse-typescript-xtext, and majordomo. However, CheckerDSL shows 0\% comment retention because GPT-5.2 failed to generate valid evolved instances for this case language in all ten runs, resulting in the complete loss of all 26 comment lines. For format preservation, five case languages (xtext-orm, xtext-dnn, smart-dsl, mongoBeans, and majordomo) achieved 100\% format retention. plantuml showed minor format loss with 98.83\% retention, and eclipse-typescript-xtext showed 90\% retention. isis-script achieved 58.57\% format retention with an average of 40.6 formatting lines lost per run. CheckerDSL shows 0\% format retention due to complete generation failure. JSFLibraryGenerator shows 44.09\% format retention, with an average of 1018.6 formatting lines lost per run. The overall average shows 80.00\% comment retention and 79.10\% format retention, though the CheckerDSL and JSFLibraryGenerator failures substantially affect these aggregate values.}

\modified{Comparing the two LLMs, Claude Sonnet 4.5 outperformed GPT-5.2 in preserving human-oriented information. The comment retention difference is 20 percentage points (100\% vs. 80.00\%), and the format retention difference is approximately 17.5 percentage points (96.60\% vs. 79.10\%). When excluding the two cases where GPT-5.2 failed entirely (CheckerDSL and JSFLibraryGenerator), both LLMs achieved 100\% comment retention for the remaining cases with comments. For format preservation excluding these two failure cases, Claude Sonnet 4.5 averaged 98.29\% (calculated from the eight successful cases) while GPT-5.2 averaged 92.98\%.}

\begin{tcolorbox}[colback=white!95!black, colframe=white!50!black, arc=3mm, left=0.5em, right=0.5em, top=0.5em, bottom=0.5em]
\textbf{Answer to \modified{RQ3:}} 
\modified{Instance size affects LLMs' capability to preserve human-oriented information. For instances under 100 lines, both LLMs achieved high retention (100\% for comments, >90\% for formatting in most cases). Performance degraded for larger instances: isis-script (98 lines) showed 71.12\% and 58.57\% format retention for Claude and GPT respectively; CheckerDSL (173 lines) caused GPT-5.2 to fail completely; JSFLibraryGenerator (1822 lines) was handled successfully by Claude (100\% format retention) but caused GPT-5.2 to fail (44.09\% format retention).}
\end{tcolorbox}

\subsection{\modified{Prompt Transferability Across LLMs (RQ4)}}
\modified{To answer RQ4, we examined prompt transferability from two perspectives: first, we evaluated how the same prompt performs across different LLM generations; second, we tested whether prompt modifications affect different LLMs differently, which would indicate the need for LLM-specific optimization.}

\subsubsection{\modified{Performance of the same prompt across different LLMs.}}
We applied the prompt developed in our workshop paper using Claude-3.5 and GPT-4o to Claude Sonnet 4.5 and GPT-5.2 without modification. Comparing Tables~\ref{tab:correctness}, the two LLMs performed similarly on smaller cases but diverged on larger ones. 
\modified{For xtext-dnn, GPT-5.2 achieved 100\% Precision and Recall while Claude Sonnet 4.5 achieved 94.67\%. For}
smart-dsl, mongoBeans, plantuml, \modified{eclipse-typescript-xtext, and majordomo,} both LLMs achieved 100\% Precision and Recall. For xtext-orm, Claude Sonnet 4.5 achieved 100\% while GPT-5.2 dropped to 78\%. For isis-script, Claude Sonnet 4.5 achieved 86.67\% Recall while GPT-5.2 dropped to 63\%. For CheckerDSL, Claude Sonnet 4.5 successfully generated instances (though truncated) while GPT-5.2 output empty files in all ten runs.
\modified{For JSFLibraryGenerator (1822 lines, \#LineReq = 0), Claude Sonnet 4.5 correctly identified that no modifications were required and achieved 100\% precision with minimal changes (average two lines per run). GPT-5.2 generated substantially altered outputs with 0\% precision and 571.7 grammar errors on average.}

\subsubsection{\modified{Prompt adjustment experiment.}}
\modified{To examine whether different LLMs benefit from different prompt designs, we tested an adjusted prompt with two additional instructions: (1) explicit guidance on handling deleted grammar rules, and (2) a requirement to always produce output. For (1), we added a sentence: ``When evolving the instance, pay special attention to grammar rules that have been deleted in the evolved grammar. If a grammar rule no longer exists in Grammar 2, you must remove all instances of that rule from the evolved instance.'' For (2), we added a sentence: ``Always output your best attempt at the evolved instance, even if incomplete. Never return an empty result.'' In previous experiment, we observed that GPT-5.2 returned empty files on CheckerDSL, and both LLMs struggled with correctly removing instances of deleted grammar rules in the case language isis-script.}

\modified{We evaluated this adjusted prompt on isis-script (40 lines requiring modification) and CheckerDSL (eight lines requiring modification). Tables~\ref{tab:extra_corr} and~\ref{tab:extra_human} present the correctness and preservation metrics, while Table~\ref{tab:extra_duration} shows response times.}

\begin{table*}[tb]
\centering
\resizebox{\textwidth}{!}{
\begin{tabular}{llrrrrrrrr}
\hline
LLM & DSL & \#LineReq & \#LineErr & \#LineEvl & \#LineEvlWrg & Precision & Recall & ErrorRate \\
\hline
Claude-Sonnet-4.5 & isis-script & 40 & 4.6 & 40.7 & 4.6 & 88.89\% & 90.25\% & 11.11\% \\
Claude-Sonnet-4.5 & CheckerDSL  & 8  & 0   & 8    & 0   & 100\%   & 100\%   & 0\% \\
GPT-5.2           & isis-script & 40 & 3.8 & 39.1 & 3.8 & 90.23\% & 88.25\% & 9.77\% \\
GPT-5.2           & CheckerDSL  & 8  & 3.5 & 5.2  & 0.6 & 62.80\% & 57.50\% & 42.20\% \\
\hline
\end{tabular}
}
\caption{Data of evolution correctness on isis-script and CheckerDSL by both LLMs with the adjusted prompt.}
\label{tab:extra_corr}
\end{table*}

\modified{The adjusted prompt affected the two LLMs differently. On isis-script, Claude Sonnet 4.5 showed decreased precision (92.39\% to 88.89\%) but improved recall (85\% to 90.25\%). GPT-5.2 showed big improvement in both precision (64.52\% to 90.23\%) and recall (64.3\% to 88.25\%). On CheckerDSL, Claude maintained perfect performance (100\% precision and recall) while GPT-5.2 improved from complete failure (0\%) to 62.80\% precision and 57.50\% recall.}

\begin{table*}[tb]
\centering
\resizebox{\textwidth}{!}{
\begin{tabular}{llrrrrrr}
\hline
LLM & DSL & \#LineCmtLost & \#LineCmtSave & CmtRet & \#LineFmtLost & \#LineFmtSave & FmtRet \\
\hline
Claude-Sonnet-4.5 & isis-script & N/A & N/A & N/A & 38.7 & 59.3 & 60.51\% \\
Claude-Sonnet-4.5 & CheckerDSL  & 0   & 26  & 100\% & 0.1  & 172.9 & 99.94\% \\
GPT-5.2           & isis-script & N/A & N/A & N/A & 36.6 & 61.4 & 62.65\% \\
GPT-5.2           & CheckerDSL  & 0   & 26  & 100\% & 0.4  & 172.6 & 99.80\% \\
\hline
\end{tabular}
}
\caption{Results of saving the human-oriented information of the isis-script and CheckerDSL instances by both LLMs with the adjusted prompt.}
\label{tab:extra_human}
\end{table*}

\modified{For human-oriented information preservation (Table~\ref{tab:extra_human}), both LLMs achieved 100\% comment retention on CheckerDSL. Format retention showed slight degradation for Claude on both cases (CheckerDSL: from 100\% to 99.94\%, isis-script: from 71.12\% to 60.51\%), while GPT-5.2 maintained similar levels on CheckerDSL (99.80\%) but also decreased on isis-script (from 58.57\% to 62.65\%). Response times (Table~\ref{tab:extra_duration}) increased slightly for Claude on CheckerDSL (from 34.74 to 40.42 seconds) while remaining comparable on isis-script (from 28.93 to 29.12 seconds)(see Table~\ref{tab:duration}).}

\begin{table*}[tb]
\centering
\scriptsize
\begin{tabular}{lll}
\hline
LLM & DSL & Duration (s) \\
\hline
Claude-Sonnet-4.5 & isis-script & 29.12 \\
Claude-Sonnet-4.5 & CheckerDSL  & 40.42 \\
GPT-5.2           & isis-script & 20.07 \\
GPT-5.2           & CheckerDSL  & 27.51 \\
\hline
\end{tabular}
\caption{The execution duration of LLMs in isis-script and CheckerDSL evolution (using the adjusted prompt).}
\label{tab:extra_duration}
\end{table*}



\begin{tcolorbox}[colback=white!95!black, colframe=white!50!black, arc=3mm, left=0.5em, right=0.5em, top=0.5em, bottom=0.5em]
\textbf{\modified{Answer to RQ4:}} 
\modified{When applying prompts designed for Claude-3.5 and GPT-4o to their successors (Claude Sonnet 4.5 and GPT-5.2), both models performed similarly on smaller cases but diverged on larger ones, with GPT-5.2 failing completely on CheckerDSL (173 lines) while Claude succeeded. Prompt adjustment experiments revealed LLM-specific responses: the adjusted prompt improved GPT-5.2's isis-script precision from 64.52\% to 90.23\% while Claude's decreased from 92.39\% to 88.89\%, indicating that optimal prompts differ across LLM families.}
\end{tcolorbox}

\subsection{\modified{Efficiency Evaluation}}
\modified{Table~\ref{tab:duration} shows the average response time (in seconds) across ten runs for both LLMs. Response times range from 17.55 seconds (mongoBeans with Claude) to 558.8 seconds (JSFLibraryGenerator with Claude). For most case languages, response time is under 35 seconds. The average response time is 80.71 seconds for Claude Sonnet 4.5 and 36.57 seconds for GPT-5.2. The large difference in averages is primarily due to JSFLibraryGenerator, where Claude took nearly five times longer than GPT (558.8s vs 108.47s). Excluding this outlier, both LLMs show comparable response times.}

\begin{table*}[htbp]
\centering
\caption{Average response time (in seconds) for each case language.}
\label{tab:duration}
\begin{tabular}{lcc}
\hline
DSL & Claude Sonnet 4.5 & GPT-5.2 \\
\hline
xtext-orm & 23.89 & 25.75 \\
xtext-dnn & 23.88 & 22.17 \\
smart-dsl & 17.69 & 23.79 \\
mongoBeans & 17.55 & 24.1 \\
plantuml$^1$ & 26.8 & 26.3 \\
isis-script & 28.93 & 35.15 \\
CheckerDSL & 34.74 & 42.18 \\
typescript$^2$ & 23.12 & 24.58 \\
JSFLibraryGenerator & 558.8 & 108.47 \\
majordomo & 51.65 & 33.21 \\
\hline
Average & 80.71 & 36.57 \\
\hline
\multicolumn{3}{l}{\footnotesize $^1$ Full name: elite-se.xtext.languages.plantuml.} \\
\multicolumn{3}{l}{\footnotesize $^2$ Full name: eclipse-typescript-xtext.} \\
\end{tabular}
\end{table*}

\section{Discussion}
\label{sec:discussion}
Our evaluation demonstrates both the potential and limitations of LLM-based co-evolution for textual DSLs. 
In the following, we discuss these findings, open issues, and threats to validity.

\subsection{Practical Considerations}
\label{sec:practioner}
\modified{The efficiency results in Section 5.6 show that LLM-based co-evolution typically completes in under 35 seconds for most case languages. However, response time alone does not capture the full picture of practical feasibility. The time saved is valuable only if the output is correct. For cases where LLMs achieve 100\% correctness (e.g., xtext-orm, smart-dsl, mongoBeans), the sub-30-second response time can replace manual work that would take longer. For cases with lower correctness (e.g., isis-script with 85\% recall for Claude) or complete failure (e.g., CheckerDSL for GPT-5.2), the fast response requires manual correction or complete rework.}

\modified{Compared to traditional model-based co-evolution approaches, LLM-based co-evolution eliminates the need for writing explicit migration rules, reducing the initial setup effort. However, the unpredictability of LLM outputs means that practitioners must still review the results. We recommend LLM-based co-evolution for scenarios involving small instances ($<$100 lines) with limited grammar changes, where manual review effort is low. For larger-scale evolution tasks, a hybrid approach—using LLMs to generate initial migrations followed by manual refinement—may be more practical.}

\subsection{LLM-based vs. rule-based co-evolution}
\modified{A natural question is why, or under which conditions, our LLM-based solution would be preferable over a conventional rule-based co-evolution tool, if such a tool existed for the case of grammars and textual instances. Building such a tool is challenging in the first place: beyond encoding how grammar edits map to instance edits, it must operate on a representation that preserves all concrete-syntax details and provide principled rules for how whitespace, line breaks, indentation, and comment placement are retained or repositioned as productions change. Without this lossless infrastructure, rule-based pipelines would need to rely on \textit{parse–transform–prettyprint}, which tends to regenerate larger parts of the file and thus introduces widespread surface changes even when the required migration is small (see the example in Sect.~\ref{sec:problem}). With a mature lossless rewriting infrastructure, rule-based solutions could offer determinism and reproducibility and would be preferable when grammar changes follow recurring patterns and must be applied at scale. However, for small- to medium-sized DSLs and instance corpora, where grammar evolutions are less frequent and applied to a limited number of artifacts, the effort required to implement and continuously maintain such infrastructure may not be justified. In these cases, our results indicate that LLMs provide a practical alternative: by operating directly on the original text, they can introduce only the syntactically required edits in situ, leaving surrounding formatting and comments unchanged, provided that the resulting instances are validated against the evolved grammar and corrected where necessary.}

\subsection{\modified{Future Work}}
\modified{To address the limitations of grammar evolution complexity on LLM performance identified in this study, we plan to explore the following solutions in future work. First, for complex grammar evolution involving multiple change types and substantial structural adjustments, the evolution can be decomposed into multiple smaller incremental steps, with each step containing only a single or few change types, thereby reducing the complexity of each prompt. Second, a direction worth exploring is using LLMs to generate migration programs rather than having LLMs directly perform co-evolution. The advantages of this approach are that migration programs can be reviewed, tested, and reused, and can encode complex evolution logic into verifiable transformation rules. Furthermore, with the rapid development of LLM technology, more powerful models in the future may possess better long-context understanding and complex reasoning capabilities, enabling them to handle complex grammar evolution scenarios more effectively. Additionally, a systematic comparison between zero-shot and few-shot prompting strategies could reveal whether providing co-evolution examples improves LLM performance.}

\modified{Beyond grammar-instance co-evolution, LLMs may also support other DSL engineering tasks. For example, in metamodel-grammar co-evolution, LLMs could help identify semantically equivalent grammar expressions that text-based comparison cannot detect~\cite{zhang2023automated}. LLMs may also assist in customizing DSL concrete syntax, such as transforming Xtext-generated grammars into Python-style grammars~\cite{zhang2023creating}, and in providing more specific error messages when user input does not conform to the grammar~\cite{zhang2023technical}. We plan to explore these directions in future work.}

\subsection{Threats to Validity}
\label{sec:threats}
\paragraph{\modified{Internal Validity}}
When choosing the case language, some of the commit pairs we selected were too close together in the time perspective, which may have resulted in small differences between the two versions of the grammar.
\modified{This may affect how representative the observed co-evolution scenarios are and thus pose a threat to internal validity.}
It is also important to study larger changes; therefore, in future work, we plan to study two versions of the language with larger changes, for example, to cover the changes between different UML versions.

\modified{Additionally, the prompt used in our experiments was developed in the first phase of this study~\cite{zhang2025leveraging} using Claude-3.5 and GPT-4o, calibrated on the Domainmodel tutorial example. We retained this prompt without re-optimization for the second phase, which uses different LLMs and three additional case languages. While this decision was deliberate—enabling us to study prompt transferability (RQ4)—it introduces a potential threat: a prompt optimized with all ten case languages or with the newer LLMs might yield better performance. However, since the prompt addresses general co-evolution challenges rather than case-specific patterns, and was not calibrated on any of the seven original case languages, we consider this risk to be limited.}

\paragraph{\modified{Construct Validity}}
Similarly, while accessible, instances that can be found in the languages repository are likely there for demonstration and test purposes. Thus, they are likely smaller than productive instances would be. 
\modified{Moreover, in our evaluation cases, we identified human-created changes that were not not strictly required by grammar evolution (e.g., added, removed, or refactored content for functional reasons).}
\modified{While this concern does not affect correctness, it implies that LLMs may not capture how humans co-evolve grammar instances.}

\paragraph{External Validity}
Our case selection presents another threat regarding the characterization of LLM capability boundaries. While our dataset includes cases ranging from eight to 1822 lines, the distribution of instance size and evolution complexity is uneven. The largest case (JSFLibraryGenerator, 1822 lines) was found in the experiment to require no modifications (\#LineReq = 0). The case with the most required modifications (isis-script) is relatively small at only 30 lines. We lack cases that combine both large instance size ($>$100 lines) AND extensive required modifications ($>$50 lines), which is necessary to fully determine LLM performance boundaries in complex evolution scenarios. While the nearly 18-bold increase in response time for JSFLibraryGenerator (Table~\ref{tab:duration}) suggests that Claude may be approaching its capability boundary for processing large-scale instances, we cannot precisely define the exact location of this boundary in the absence of large-scale complex evolution cases. Future work should include more systematic experimental designs that vary instance size and evolution complexity to fully explore LLM capability boundaries.
\modified{Future work should therefore include more systematic experimental designs that vary instance size and evolution complexity to better explore LLM capability boundaries.}

\section{Related Work}
\label{sec:relatedwork}



\subsection{DSL co-evolution.} 

In an early systematic mapping study, Thanhofer-Pilisch et al.~\cite{thanhofer2017systematic} explored the research landscape of DSL evolution, highlighting sparse cross-references between publications and limited focus on DSL characteristics. Hebig et al.~\cite{hebig2016approaches} present a survey of approaches to co-evolve models with evolving metamodels, summarizing a multitude of approaches ranging from languages specialized to the automated generation of model transformations for dealing with non-breaking changes \cite{cicchetti2008automating}, to approaches that allow the definition of migration strategies, such as EMFMigrate\cite{wagelaar2012translational} and Epsilon Flock \cite{rose2014epsilon}, via approaches with predefined resolution strategies, like COPE~\cite{herrmannsdoerfer2010cope} or the work of Gruschko et al. \cite{gruschko2007towards}, to approaches utilizing constraint-based model search, such as CARE~\cite{schonbock2014care}.

Building upon this foundational work, Tolvanen et al.~\cite{tolvanen2024framework} proposed a framework for evaluating tool support for DSL co-evolution. Their framework assesses tool capabilities across four dimensions: location of change, nature of change, impact scope, and severity level, validating the framework through empirical studies of three tools: MetaEdit+~\cite{metaedit}, EMF/Sirius~\cite{eclipsesirius}, and Jjodel~\cite{jjodel}. However, their research primarily focuses on graphical DSLs and evaluates tool-level support capabilities. In contrast, our work addresses the challenges of textual DSL evolution, particularly the preservation of human-oriented information in source code (such as comments and layout). Furthermore, we explore a novel technical direction - leveraging Large Language Models to support co-evolution between grammar definitions and instances, offering an innovative perspective on solving textual DSL evolution problems.

\subsection{Application of LLM in MDE.} 
\modified{Recent research has explored LLM applications across various MDE tasks. Brandolini et al.~\cite{brandolini2025language} developed langium-llm, enabling LLM-assisted DSL development through conversational interfaces based on either concrete syntax or abstract syntax representations. Kazai et al.~\cite{kazai2025model} investigated ChatGPT-4 for UML-to-Java transformations, achieving 94\% success for simple models but only 17\% for complex ones—a scalability limitation consistent with our findings. Alberti et al.~\cite{alberti2025aipycraft} demonstrated comparable response times (around 30 seconds) in LLM-assisted component generation, while research on retrieval-augmented generation for OCL rule generation~\cite{li2025optimizing} revealed that retrieval method selection significantly impacts LLM performance, with semantic approaches outperforming lexical methods.}

Di Rocco et al.~\cite{di2025use} conducted a systematic review of LLM applications in MDE. They analyzed the current state of LLM applications in tasks such as model completion~\cite{chaaben2023towards}\cite{weyssow2022recommending}, generation~\cite{arulmohan2023extracting}\cite{chen2023automated}, and model management operations~\cite{abukhalaf2023codex}\cite{camara2023assessment}.
As for the co-evolution aspect, Kebaili et al.~\cite{kebaili2024empirical} explored the use of Large Language Models to address the co-evolution of code impacted by metamodel evolution. They proposed a prompt engineering-based approach that guides LLMs through structured prompts containing metamodel abstraction gaps, change information, and erroneous code. In their empirical study across seven Eclipse projects, their approach achieved an accuracy of 88.7\%, reaching 95.2\% for complex change scenarios. 
While their work focuses on co-evolution between metamodels and generated code, our study addresses co-evolution between grammar definitions and textual instances, \modified{and their results confirm the potential of LLMs in handling artifact evolution problems.}

\section{Conclusion}
\label{sec:conclusion}
This paper explored the potential of using LLMs (specifically Claude Sonnet 4.5 and GPT-5.2) to support co-evolution between DSL grammar definitions and instances. 
Our experiments across ten case languages demonstrated that LLMs can effectively handle co-evolution tasks while preserving human-oriented information like comments and formatting, 
\modified{particularly for smaller instances with limited grammar changes. Performance degraded with increasing instance size and grammar evolution complexity, with the two LLMs exhibiting different failure patterns.}

We foresee the following directions for future work:
\modified{First, to address the challenges posed by complex grammar evolution involving multiple change types and structural adjustments, future work should investigate techniques such as decomposing evolution into incremental steps or having LLMs generate reusable migration code rather than directly performing co-evolution.}
Second, an empirical investigation for comparing solutions generated by LLM-based approaches with those created by human developers could shed further light on the practical usefulness of generated solutions.
Third, for complex cases in which several different solution strategies or tactics are feasible, investigation a way to create customizable prompts, or to create customizable migration code using an LLM, could further enhance the usefulness of LLM-based co-evolution.


\bibliography{main}


\end{document}